# A CRITICAL FIELD GUIDE FOR WORKING WITH MACHINE LEARNING DATASETS

Written by Sarah Ciston {1}
Editors: Mike Ananny {2} and Kate Crawford {3}

Part of the <u>Knowing Machines</u> research project

# TABLE OF CONTENTS

# INTRODUCTION TO MACHINE LEARNING DATASETS

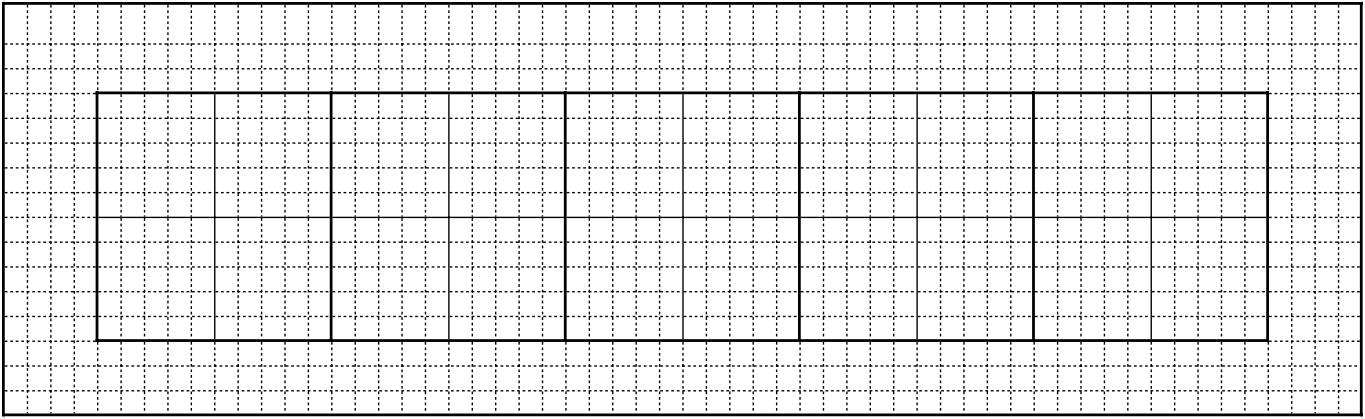

Maybe you're an engineer creating a new machine vision system to track birds. You might be a journalist using social media data to research Costa Rican households. You could be a researcher who stumbled upon your university's archive of handwritten census cards from 1939. Or a designer creating a chatbot that relies on large language models like GPT-3. Perhaps you're an artist experimenting with visual style combinations using DALLE-2. Or maybe you're an activist with an urgent story that needs telling, and you're searching for the right dataset to tell it.

No matter what kind of datasets you're using or want to use, whether you're curious but intimidated by machine learning or already comfortable, this work is complicated. Because machine learning relies on datasets, and because datasets are always tangled up in the ways they're created and used, things can get messy. You may have questions like:

How do the dataset pre-processing methods I choose affect my outcomes?

How might this dataset contribute to creating errors or causing harm?

More than likely you will encounter at least some of these conundrums — as many of us who work with machine learning datasets do. Anyone using datasets will weigh choices and make tradeoffs. There are no universal answers and no perfect actions — just a tangle of dataset forms, formats, relationships, behaviors, histories, intentions, and contexts.

When choosing and using machine learning datasets, how do you deal with the issues they bring? How can you navigate the mess thoughtfully and intentionally? Let's jump in.

INTRODUCTION TO MACHINE LEARNING DATASETS

1.1

# WHAT IS THIS GUIDE ?

Machine learning datasets are powerful but unwieldy. They are often far too large to check all the data manually, to look for inaccurate labels, dehumanizing images, or other widespread issues. Despite the fact that datasets commonly contain problematic material — whether from a technical, legal, or ethical perspective — datasets are also valuable resources when handled carefully and critically. This guide offers questions, suggestions, strategies, and resources to help people work with existing machine learning datasets at every phase of their lifecycle. Equipped with this understanding, researchers and developers will be more capable of avoiding the problems unique to datasets. They will also be able to construct more reliable, robust solutions, or even explore promising new ways of thinking with machine learning datasets that are more critical and conscientious. {4}, {5}

If you aren't sure whether this guide is for you, consider the many places you might find yourself working with machine learning datasets. This guide can be helpful if you are…

- making a model

- working with a pre-trained model

- researching an existing machine learning tool

- teaching with datasets

- creating an index or inventory

- concerned about how datasets describe you or your community

- learning about datasets by exploring one

- stewarding or archiving datasets

- investigating as an artist, activist, or developer

> This list is non-exhaustive, of course. Datasets are being used widely across countless domains and industries. How else can you imagine working with machine learning datasets?

The appetite for massive datasets is huge and still accelerating, fueled by the perceived promise of machine learning to convert data into meaningful, monetizable information.[6] Too often, this work is done without regard for how datasets can be partial, imperfect, and historically skewed. Take the widely publicized examples of police departments and courts selecting "future criminals" from software that relied on historical crime records, which ProPublica journalists found was grossly inaccurate and targeted Black people in its predictions [6]. More troubling still, researchers (and public and private organizations) continue to make use of such datasets despite learning of their harms — perhaps because they seem more efficient or effective, because they are already part of common practices in their communities, or simply because they are the most readily available options. This is exactly why critical care is so needed — datasets' potential harms are subtle, localized, and complex. You will need to make conscientious decisions and compromises when working with any dataset. There is no perfect representation, no correct procedure, and no ideal dataset.

This guide aims to help you navigate the complexity of working with datasets, giving you ways to approach conundrums carefully and thoughtfully. Section 1 describes how DATA and DATASETS are dynamic research materials, and Section 2 outlines the BENEFITS of working critically with datasets. Then you'll find more on the common PARTS of datasets (Section 3), examples of the TYPES of datasets you may encounter (Section 5), and how to TRANSFORM datasets (Section 4) — all to help make critical choices easier. Then Section 6 provides a DATASET LIFECYCLE framework with important questions to ask as you engage critically at each stage of your work. Finally, Section 7 offers some CAUTIONS & REFLECTIONS for careful dataset stewardship.

## FIELD GUIDES AND DATASETS AS FORMS

> The field guide format frames this text because, like datasets, field guides teach their readers particular ways of looking at the world — for better and for worse. Carried in a knapsack, a birder might use their field guide to confirm a species sighting in a visual index. A hiker might read trail warnings to prepare for their trek. With these practical uses, the field guide speaks to a desire to connect deeply with dataset tools and practices, and a sense of careful responsibility that data stewardship shares with environmental stewardship. However, naturalist field guides also draw on the same problematic histories of classifying and organizing information that are foundational to many machine learning tasks today. This critical field guide aims to help bring understanding to the complexities of datasets, so that the decisions you make while using them are conscientious. It invites you to mess with these messy forms and to approach any logic of classification with a critical eye.

> When we say **CLASSIFICATION** in this guide, generally we refer to the choices, logics, and paradigms that inform sociotechnical communities of practice — how people sort and are sorted into categories, how those categories come to be seen as dominant and naturalized, and how people are differently affected by those categories. We acknowledge that the term **CLASSIFICATION** also refers to specific machine learning tasks that label and sort items in a dataset by discrete categories. For example, asking whether an image is a dog or a cat is handled by a classification task. These are distinguished from **REGRESSION** tasks, which show the relationship between features in a dataset, for example sorting dogs by their age and number of spots. In this guide, we will specify 'tasks' when referring to these techniques, but simply say 'classification' when referring to the sociotechnical phenomenon more broadly.{7}

# WHAT ARE DATA?

### DATA ARE CONTINGENT ON HOW WE USE THEM

DATA are values assigned to any 'thing', and the term can be applied to almost anything. Numbers, of course, can be data; but so can emails, a collection of scanned manuscripts, the steps you walked to the train, the pose of a dancer, or the breach of a whale. *How* you think about the information is what makes it data. Philosopher of science Sabina Leonelli sees data as a "relational category" meaning that, "What counts as data depends on who uses them, how, and for which purposes." She argues data are "any product of research activities [...] that is collected, stored, and disseminated *in order to be used as evidence for knowledge claims*" [8]. This definition reframes data as contingent on the people who create, use, and interact with them in context.

### DATA MUST BE MADE, AND MAKING DATA SHAPES DATA

As a form of information,{8} data do not just exist but have to be generated, through collection by sensors or human effort. Sensing, observing, and collecting are all acts of interpretation that have contexts, which shape the data. For example, when collecting images of faces using infrared cameras, that data can provide heat signatures but not the eye color of its subjects. Studies are designed with specific equipment to achieve their goals and not others. Whether they are quantitative data captured with a sensor or qualitative data described in an interview, the context in which that data is collected has already created a limit for what it can represent and how it can be used. It is easy to think that calling information "data" makes it discrete, separate, fixed, organized, computable — static [1]. But dataset users impose these qualities on information *temporarily* — to organize it into familiar forms that suit machine learning tasks and other algorithmic systems.

MACHINE LEARNING, DEEP LEARNING, NEURAL NET, ALGORITHM, MODEL —
WHAT'S THE DIFFERENCE?

An **ALGORITHM** is a set of instructions for a procedure, whether in the context of machine learning or another task. Algorithms are often written in code for machines to process, but they are also widely used in any system of step-by-step instructions (e.g. cooking recipes). Algorithms are not a modern Western invention, but predate computation by thousands of years, as technology culture researcher Ted Striphas has shown [16]. That said, algorithms stayed associated mainly with mathematical calculation until quite recently, according to historian of science Lorraine Datson, who traces their expansion into a computational catch-all in the mid-20th century [17].

A **MODEL** is the result of a machine learning algorithm, once it includes revisions that take into account the data it was exposed to during its training. It is the saved output of the training process, ready to make predictions about new data. One way to think of a model is as a very complex mathematical formula containing millions or billions of variables (values that can change). These variables, also called model parameters, are designed to transform a numerical input into the desired outputs. The process of model training entails adjusting the variables that make up the formula until its output matches the desired output.

Much focus is put on machine learning models, but models depend directly on datasets for their predictions. While a model is not part of a dataset, it is deeply shaped by the datasets it is based upon. Traces of those datasets remain embedded within the model no matter how it is used next. (This guide won't cover the detailed aspects of working critically with machine learning models and understanding how they learn — that's a whole other discussion. Terms like 'activation functions', 'loss functions', 'learning rates', and 'fine-tuning' give a taste of the many human-guided processes behind model making, an active conversation and set of practices that are beyond the scope of this guide.)

Artificial **NEURAL NETWORKS** describe some of the ways to structure machine learning models (see TYPES in Section 4), including making large language models. Named for the inspiration they take from brain neurons (very simplified), they move information through a series of nodes (steps) organized in layers or sets. Each node receives the output of the previous layers' nodes, combines them using a mathematical formula, then passes the output to the next layer of nodes.

> **MACHINE LEARNING** is a set of tools used by computer programmers to find a formula that best describes (or models) a dataset. Whereas in other kinds of software the programmer will write explicit instructions for every part of a task, in machine learning, programmers will instruct the software to adjust its code based on the data it processes, thus "learning" from new information [18]. Its learning is unlike human understanding and the term is used metaphorically.
>
> Some formulas are "deeper" than others, so called because they contain many more variables, and **DEEP LEARNING** refers to the use of complex, many layers in a machine learning model. Due to their increasing complexity, the outputs of machine learning models are not reliable for making decisions about people, especially in highly consequential cases. When working with datasets, include machine learning as one suite of options in a broader toolkit — rather than a generalizable multi-tool for every task.

# WHAT ARE DATASETS?

A DATASET can be any kind of collected, curated, interrelated data. Often, datasets refer to large collections of data used in computation, and especially in machine learning. Information collections are transformed into datasets through a LIFECYCLE of processes (collection/selection, cleaning and analyzing, sharing and deprecating), which shape how that information is understood. (*For critical questions you can ask at each phase of a dataset's lifecycle, see Section 6.*)

## DATASETS ARE TIED TO THEIR MAKERS

The many choices that go into dataset creation and use make them extremely dynamic. They always reflect the circumstances of their making — the constraints of tools, who wields them and how, even who can afford the equipment to train, store, and transmit data. For example, analysis of national datasets in the Processing Citizenship project revealed how some European nations collected information differently, with a range of specificity in categories like 'education level' or 'marital status'. Software engineer Wouter Van Rossem and science and technology studies professor Annalisa Pelizza examined not only the data in that dataset, but how they were labeled, organized, and utilized to show that these reflected how nations perceived the migrants they cataloged [19]. When gathered by a different group, using different tools, a dataset will be quite different — even if it attempts to collect similar information.

### DATASETS ARE TIED TO THEIR SETTINGS

Datasets can be frustratingly limited, but this does not mean they are static; instead, the information in datasets is always wrapped up in the contexts that make and use them. Media scholar Yanni Alexander Loukissas, author of All Data Are Local, calls datasets "data settings," arguing that "data are indexes to local knowledge." They remain tied to the communities, individuals, organisms, and environments where they were created. Instead of treating data as independent authorities, he says we should ask, "*Where* do data direct us, and *who* might help us understand their origins as well as their sites of potential impact?" [1]. These questions extend the possibilities for exploring datasets as dynamic materials. Therefore, datasets must be used carefully, with consideration for their material connection to their origins.

**For your consideration:** How does framing information as "data" change your relationship to it? What other forms of information do you work with? What kinds of information should not be included in datasets?

# BENEFITS: WHY APPROACH DATASETS CRITICALLY?

HERE ARE SOME EXAMPLES OF HOW DATASET STEWARDSHIP CAN BENEFIT YOUR PRACTICE, AS WELL AS BENEFIT OTHERS:

**MORE ROBUST DATASETS** ARISE BY CONSIDERING MULTIPLE PERSPECTIVES AND WORKING TO REDUCE BIAS.

> TO-DO: Include interdisciplinary, intersectional communities in designing, developing, implementing, and evaluating your work. (See ALTERNATIVE APPROACHES TO DATASET PRACTICES)

**MORE RELIABLE RESULTS** COME FROM ANTICIPATING AND ADDRESSING CONTINGENCIES LIKE DEPRECATED DATASETS AND UNINFORMED CONSENT.

> TO-DO: Apply checkpoints at each stage, asking critical questions about data provenance and reflecting on your own methodologies.

**GAIN INCREASED PROTECTION FROM LIABILITY** FOR DATASETS WITH LEGAL OR ETHICAL ISSUES BY PROACTIVELY ADDRESSING POTENTIAL CONCERNS BEFORE USE.

> TO-DO: This does not constitute legal advice. However, always perform due diligence before working with existing datasets, including checking any licenses or terms of use. Simply downloading some datasets can create legal liability [20], [21]. So try to be aware of potential consent issues, misuse, or ethical concerns beyond those outlined by the dataset creators, especially as they may have changed since creation or arise from your new usage. You can check data repositories and data journalism to see how datasets have already been used.

**CRITICAL PRACTICES ARE BECOMING FIELD-STANDARD** AND REQUIRED FOR ACCESS TO TOP CONFERENCES AND JOURNALS.

> TO-DO: Help shape the future of the field by modeling and advocating for best practices. Suggest new frameworks and methods for making, using, and deprecating datasets.

**MORE CAREFUL AND CONSCIENTIOUS OUTCOMES** FOR THOSE IMPACTED BY RESULTS.

> TO-DO: Engage the people and groups affected by datasets and your use of them, to learn what careful and conscientious practices mean to them.

**OPEN-SOURCE, OPEN-ACCESS, AND OPEN RESEARCH COMMUNITIES** BUILD POSITIVE FEEDBACK LOOPS THROUGH DATASET STEWARDSHIP OF RELIABLE MATERIALS.

> TO-DO: Share datasets responsibly, through centralized repositories and with thorough documentation.

**NO NEUTRAL CHOICE (OR NON-CHOICE) EXISTS.** "WHEN THE FIELD OF AI BELIEVES IT IS NEUTRAL," SAYS AI RESEARCHER PRATYUSHA KALLURI, IT "BUILDS SYSTEMS THAT SANCTIFY THE STATUS QUO AND ADVANCE THE INTERESTS OF THE POWERFUL" [23].

> TO-DO: Working with datasets brings challenges that need conversations and multiple perspectives. Discuss issues with your team using the Dataset's Lifecycle questions in Section 6, plus the wide range of critical positions shared in the "Critical Dataset Studies Reading List" compiled by the Knowing Machines research project [22].
>
> Before publishing or launching your work, ask hard questions and share your project with informal readers within your networks who can provide constructive feedback. Go slow. Pause or even stop a project if needed.
>
> Remember that taking "no position" on a dataset's ethical questions is still taking a position. Consider the tradeoffs for choosing one dataset or technique over another.

# PARTS OF A DATASET

What actually makes up a machine learning dataset, practically speaking? Here are some of the key terms that are helpful for understanding their parts and dynamics:

## INSTANCE

One data point being processed or sorted, often viewed as a row in a table. For example, in a training dataset for a classification task that will sort images of dogs from cats, one instance might include the image of a dog and the label "dog," while another instance would be an image of a cat and the label "cat" as well as other pertinent metadata (see also LABEL, METADATA, and TRAINING DATA below in this section and SUPERVISED machine learning in Section 4).

## FEATURE

One attribute being analyzed, considered, or explored across the dataset, often viewed as a column in a table. Features can be any machine-readable (i.e. numeric) form of an instance: images converted into a sequence of pixels, for example. Note: Researchers often select and "extract" the features most relevant for their purpose. Features are not given by default. They are the results of decisions made by datasets' creators and users. (For more discussion of ENGINEERING FEATURES see Section 5.)

### LABEL

The results or output assigned by a machine learning model, or a descriptor included in a training dataset meant for the model to practice on as it is built, or in a testing or benchmark dataset used for evaluation or verification. (See Section 7.2.2 for more on labels' creation and their potentially harmful impacts.)

### METADATA

Data about data, metadata is supplementary information that describes a file or accompanies other content, e.g. an image from your camera comes with the date and location it was shot, lens aperture, and shutter speed. Metadata can describe attributes of content and can also include who created it, with what tools, when, and how. Metadata may appear as TABULAR DATA (a table) and can include captions, file names and sizes, catalog index numbers, or almost anything else. Metadata are often subject- or domain-specific, reflecting how a group organizes, standardizes and represents information [23].

### DATASHEET

A document describing a dataset's characteristics and composition, motivation and collection processes, recommended usage and ethical considerations, and any other information to help people choose the best dataset for their task. Datasheets were proposed by diversity advocate and computer scientist Timnit Gebru, et al., as a field-wide practice to "encourage reflection on the process of creating, distributing, and maintaining a dataset, including any underlying assumptions, potential risks or harms, and implications for use" [24]. Datasheets are also resources to help people select and adapt datasets for new contexts.

### SAMPLE

A selection of the total dataset, whether chosen at random or using a particular feature or property; samples can be used to analyze a dataset, perform testing, or train a model. For more on practices like sampling that transform datasets, see Section 5.

### TRAINING DATA

A portion of the full dataset used to create a machine learning model, which will be kept out of later testing phases. Imagining a model like a student studying for exams, you could liken the training data to their study guide which they use to practice the material. For example, in supervised machine learning (see Section 4), training data includes results like those the model will be asked to generate, e.g. labeled images. Training datasets can never be neutral, and they commonly "inherit learned logic from earlier examples and then give rise to subsequent ones," says critical AI scholar Kate Crawford [25].

### VALIDATION DATA

A portion of the full dataset that is separated from training data and testing data, validation data is held back and used to compare the performance of different design details. Validation data is separate from testing data, because validation data is used during the training process to optimize the model while adjustments are being made; therefore, the resulting model will be familiar with its data. That means separate testing data is still needed to confirm how the final model performs. Imagine validation data as practice tests that programmers can administer to check on the model's progress so far.

### TESTING DATA

A portion of the full dataset that is separated from the training data and validation data, and that is not involved in creation of a machine learning model. Testing data is then run through the completed model in order to assess how well it functions. Testing data for the model would be similar to the student's final exam.

### TENSORS: SCALARS, VECTORS, MATRICES (oh my!)

Software for working with machine learning datasets organizes information in numerical relationships, in grids called TENSORS. Understanding tensors can help you understand how data are viewed, compared, and manipulated in computational models. Their grids can have many dimensions, not only two-dimensional X-and-Y graphs [26]. A SCALAR describes a single number. A VECTOR is a list (aka an array), like a line of numbers. A MATRIX is a 2D tensor, like a rectangle of numbers. And a grid of three (or more) dimensions is a TENSOR, like a cube of numbers, or a many-dimensional cube of numbers.

### DATA SUBJECTS

The people and other beings whose data are gathered into a dataset. Even if identifying information has been removed, datasets are still connected to the subjects they claim to represent.

### DATA SUBJECTEES

This new and somewhat unwieldy term is used here to describe people impacted directly or indirectly by datasets, distinct from data subjects. Data subjectees include anyone affected by predictions made with machine learning models, for example someone forced to use a facial detection system to board a flight or eye-tracking software to take a test at school. Similarly, Batya Friedman and David G. Hendry of the Value Sensitive Design Lab distinguish between "direct" and "indirect stakeholders" to describe the different types of entanglement with technologies [27].

**For your consideration:** What other parts of a dataset are not included here but could be? How do you see dataset parts differently when you consider them within "data settings," or contexts, tied to data subjects and data subjectees?[1] What kinds of contexts are impossible to include in datasets?

# TYPES OF DATASETS

# WHAT DISTINGUISHES TYPES OF DATASETS?

You may choose a dataset based on what it contains, how it is formatted, or other needs. For example, computer vision datasets include thousands of **IMAGE** or **VIDEO** files, while natural language processing datasets contain millions of bytes of **TEXT**. You may work with waveforms as **SOUND** files or time series data, or network **GRAPH** stored in structured text formats like **JSON**. In tables you might find **PLACE** data as geographic coordinates or X-Y-Z coordinates, **TIME** as historical date sequences or milliseconds. Likely, you'll work with other types, too, or with combinations of **MULTIMODAL** data. Each dataset may include corresponding **METADATA**, documentation, and (hopefully) a complete **DATASHEET**.

You can also consider datasets based on whether the information is **STRUCTURED**, such as tabular data formatted in a table with labeled columns, or **UNSTRUCTURED**, such as plain text files or unannotated images. Annotating or coding a dataset prepares it for analysis, including supervised machine learning; and annotation raises important questions about labor, classification, and power. (See Section 6.1 for more on annotation and labeling.)

Datasets for **SUPERVISED** machine learning need to include labels for at least a portion of the data that the system is designed to "learn." This means, for example, that a dataset for object recognition would contain images as well as a table to describe the manually located object(s) they contain. It might have columns for the object name or label, as well as coordinates for the object position or outline, and the corresponding image's file name or index number.

In contrast, **UNSUPERVISED** machine learning looks for patterns that are not yet labeled in the dataset. It uses different kinds of machine learning algorithms, such as clustering groups of data together using features they share. However, it would be a misnomer to think that conclusions drawn from unsupervised machine learning are somehow more pure or rational. Much human judgment goes into developing an unsupervised machine learning model — from adjusting weights and parameters to comparing models' performance. Often supervised and unsupervised approaches are used in combination to ask different kinds of questions about the dataset. Other kinds of machine learning approaches (like reinforcement learning) don't fall neatly into these high-level categories.

(For a discussion of deprecated datasets, see Section 7.2.3, and for critical questions at every stage of working with datasets, see Section 6.)

# EXAMPLES: HOW HAVE RESEARCHERS, ENGINEERS, JOURNALISTS, AND ARTISTS PUT DATASETS TO USE?

When starting a project, you may not know what kind of dataset you need. You might work with a particular kind of media or file type most often, so you start there — or maybe you want to try a new form. You may start with a curiosity, and you're open to datasets in any format, from any source. To spark your imagination, here are four projects that used pre-existing datasets in novel and creative ways:

### JOURNALISTS UNCOVER RAINFOREST EXPLOITATION WITH GEOSPATIAL DATA

Brazilian investigative journalists at Armando.info, collaborating with El País and Earthrise Media, used field reports and satellite images to find deforestation, hidden runways, and illegal mining in the Venezuelan and Brazilian Amazon. Through computer vision analysis developed from analog maps and used on imagery from a European Space Agency satellite, the journalists compared this analysis with existing information, including complaints from Indigenous communities. "It's not that this was a technology-only job," says Joseph Poliszuck, Armando.info's co-founder. "The technology allowed us to go into the field without being blindfolded" [28]. Using similar methods, Pulitzer fellow Hyury Potter detected approximately 1,300 illegal runways, more than the number of legally registered ones in the Brazilian Amazon. Combining data work with fieldwork in international collaborations helped these journalists connect local stories to larger scale climate crises and to support communities' efforts to create change.

### HISTORIANS ASSEMBLE FRAGMENTS OF ANCIENT TEXTS

Researchers from Google's DeepMind used a neural net on an existing scholarly dataset to complete, date, and find attributions for fragments of ancient texts. They drew on 178,551 ancient inscriptions written on stone, pottery, metal, and more. that had been transcribed in the text archive Packard Humanities Institute's Searchable Greek Inscriptions [29]. They said that the "process required rendering the text machine-actionable [in plain text formats], normalizing epigraphic notations, reducing noise and efficiently handling all irregularities" [30]. They collaborated with historians and students to corroborate the machine learning outputs, calling it a "cooperative research aid" showing how machine learning research can include humans in the training process. They also created an open-source interface: ithaca.deepmind.com

### SOUND ARTIST EXPERIMENTS WITH REFUGEE ACCENT DETECTION TOOLS

Pedro Oliveira's work [31] explores the accent recognition software used since 2017 by the German Federal Office for Migration and Refugees (BAMF). Though BAMF does not disclose the software's datasets, Oliveira traced the probable source to two annotated sound databases from the University of Pennsylvania — unscripted Arabic telephone conversations named "CALL FRIEND" [32] and "CALL HOME" [33]. In 2019 the software had an error rate of 20 percent despite its deployment 9,883 times in asylum seekers' cases [34]. Oliveira utilizes sounds removed from the datasets and reverse engineers the algorithm (as musical transformations rather than for classification tasks), in order to show how politically charged it is to define and detect accents. "How can you say it's an accurate depiction of an accent?" he says. "Arabic is such a mutating language. That's the beauty of it actually" [35]. He presents this through live performance and the online sound essay "On the Apparently Meaningless Texture of Noise" [36].

### HUMAN RIGHTS ACTIVISTS ACCOUNT FOR WAR CRIMES WITH SYNTHETIC DATA

Sometimes important training data is missing from a dataset, because not enough of it exists, and these absences can amplify narrow assumptions about a diverse community. In other cases that don't involve human subjects, synthetic data can fill gaps in creative ways. When human rights activists from Mnemonic, who were investigating Syrian war crimes using machine learning, struggled to find enough images of cluster munitions to train their model, computer vision group VFRAME created synthetic data — 10,000 computer-generated 3D images of the specialized weapon and its blast sites — which researchers then used to sift through the Syrian Archive's 350,000 hours of video, searching for evidence of war crimes [37], [38]. Such systems can reduce the number of videos people need to comb through manually, while still keeping humans involved with pattern review and confirmation.

THERE ARE MANY MORE EXAMPLES LIKE THESE OF HOW TO SOURCE, USE, AND COMBINE DATASETS THAT ALREADY EXIST. THE CRITICAL AND CREATIVE POSSIBILITIES ARE NEARLY ENDLESS.

> **For your consideration:** What kind of dataset(s) will you use, and how can you approach it more critically? How will you apply what you've learned here to your next machine learning project?

# TRANSFORMING DATASETS

Just as there is no such thing as neutral data, no dataset is ready to use off the shelf. From preprocessing (sometimes confusingly called 'cleaning') to model creation, transformations reflect the perspectives of the dataset creators and users. This overview covers some of the technical details of getting a dataset ready for your tasks [18], [23], [26], [39], [40], [41], {9}, while asking critical questions along the way.

As artist and researcher Kit Kuksenok argues, "Data cleaning is unavoidable! Each round of repeated care of data and code is an opportunity to invite new perspectives to code/data technical objects" [42]. Preprocessing is a key part of building any system with a dataset, so it is crucial to document and reflect upon preprocessing transformations.

***CAUTION:*** *Be on the lookout for dataset transformations that result in lost meanings, new misconceptions, or skewed information.*

## STORING DATA

A dataset must live somewhere, and once it grows beyond a single manageable file, it usually lives in a DATABASE. While 'dataset' describes *what* the data are, 'database' describes *how* data are stored — whether as a set of tables with columns and rows (e.g. a relational database like `SQL`, or a collection of documents with keys and values like `MongoDB`). Database structures should suit what they hold, but they will also *shape* what they hold and reflect how database designers see data. "They also contain the legacies of the world in which they were designed," says media studies scholar Tara McPherson [43].

## ACCOUNTING FOR MISSING DATA

You may have entries in your dataset that read `NaN` (not a number) or `NULL`, which may or may not cause errors, depending on what kinds of calculations you do. You may also have manual entries like, '?', 'huh', or blanks that lack context. Should you remove the missing information? If you replace it, how will you know what goes in its place? Is data missing in uniform ways such that whole categories can be eliminated, or is it only missing for subgroups in ways that could skew results? How will you know what impacts your edits may have? Consider what missing data might mean. "Unavailable [is] semantically different from data that was simply never collected in the first place," says data scientist David Mertz [41]. Filtering out data and filling in data have very different implications. Could you consult data subjects to get more context on missing data or the implications of removal or substitutions? How have others handled similar challenges? Can you run tests that treat missing data differently and compare the results? Mimi Ọnụọha's "The Library of Missing Datasets" reflects on how missing data imply what will not or cannot be collected, or what has been considered not worthy of collection. The project creates a physical archive of empty files, covering topics that are excluded despite our data-hungry culture. She says, "That which we ignore reveals more than what we give our attention to" [44].

## HANDLING EXTRA DATA

You'll probably encounter dataset anomalies, outliers, and duplicates and then need ways to identify, adjust, or remove them. In text datasets for unsupervised learning, you'll likely remove punctuation and "stop words" (commonly used conjunctions or articles like 'an' or 'the', for example). But, as software engineer Francois Chollet says, "even perfectly clean and neatly labeled data can be noisy when the problem involves uncertainty and ambiguity" [18]. Outliers can also be accurate and contain meaningful information. As Crawford emphasizes, such acts of data cleaning and categorization create their own concepts of outside and otherness that can restrict "how people are understood and can represent themselves" [25]. Defining outliers, anomalies, or extra data means deciding what is 'normal', unexpected, or distracting — what is signal and what is noise.

## DISCRETIZING DATA:

AKA "binning" or grouping instances together may be useful when you don't need the original level of detail provided (see 'dimensionality reduction' in ENGINEERING FEATURES, below). For example, you might switch continuous data like temperature readings into bins grouped by every five or ten degrees. There are built-in functions for doing so, but remember that creating data ranges can skew results and may not be appropriate for all cases. Make sure to document any changes and provide the original dataset as well as the modified version.

## TOKENIZING OR CHUNKING DATA:

Breaking up data into smaller units. Tokens are often individual words or sentences. Other text-related operations might include removing punctuation and stop words that are commonly used (see HANDLING EXTRA DATA). Datasets should use tokenization strategies that account for linguistic differences, since a 'word' as a unit of meaning can vary significantly among languages. Other sequence-like data types like audio and video are also broken up into chunks, sometimes in order to be processed, compressed, or streamed.

## NORMALIZING DATA

Altering numerical data to bring them all within the same range and using the same units, e.g. between 0–1, is called normalization, or scaling [18]. In text datasets, this can also mean converting all text to lowercase and standardizing each word by reducing it to its root form (stemming or lemmatizing). In image data, this could also mean cropping images to the same dimensions or around the same subject, changing the color profile to grayscale, and so on. Remember that, while useful for many tasks, normalizing data potentially removes important context or adds ambiguity to the data (e.g. cropped images ignore any information outside the frame, acronyms may be read as their homonyms, scaled numbers may then be rounded and lose specificity).

## SEPARATING TESTING & VALIDATION DATA

If it has not already been separated, mark off a portion of your data that will not be exposed to your model or used to train it in any way. Keep these testing and validation portions separate from your training data, so that they can be used later to test your model's performance on new information once it has been trained on the remaining training data. (See Section 3, TRAINING DATA and VALIDATION DATA for more context.)

## ENGINEERING FEATURES

You may need to create features (e.g., add columns to your table) to show data from new perspectives. This can impact how the dataset can be analyzed going forward, how the model can be designed, and how the data subjects and subjectees might be affected. For example, the unsupervised machine learning approach called 'dimensionality reduction' analyzes a dataset for its most relevant features so that the rest can be ignored. It can also involve combining or altering existing features to simplify the whole. However, this runs directly counter to what legal scholar Kimberlé Crenshaw calls intersectional analysis — an approach that rejects grouping people together in categories without attending to the unique experiences (and the data) of people for whom those categories intersect, who are most negatively impacted by systems with the power to categorize people [45] (For more on intersectionality, see ALTERNATIVE APPROACHES TO DATASET PRACTICES).

### EXPLORING DATA

Sorting, sampling, combining, pivoting, and visualizing data are other transformations you will likely use during dataset preprocessing. These will differ greatly depending on the type of dataset and your project's objectives, but they all require asking critical questions about how such explorations influence the meaning of the dataset, the model developed, and the system deployed.

**For your consideration:** How have your data transformations shaped your dataset so far? Which transformations are most thought-provoking and worth exploring? Who else could offer perspective on your preprocessing decisions?

# THE DATASET'S LIFECYCLE

DATASETS OFTEN HAVE SURPRISING HISTORIES, USES, AND AFTERLIVES. FROM LOCATING THE BEST DATASET FOR THE JOB, TO WORKING WITH IMPERFECT EXISTING DATA, TO SHARING RESULTS AND ACCOUNTING FOR IMPACTS, HERE ARE SOME CRITICAL QUESTIONS TO ASK AT EACH STAGE OF A PROJECT THAT USES MACHINE LEARNING DATASETS:

# ORIGINS: WHAT IS YOUR DATASET'S STORY?

"Datasets are the results of their means of collection," says artist and technology researcher Mimi Ọnụọha [46]. They are influenced by the many people who contribute to them, who participate in their creation (knowingly or unknowingly), who collect data, who annotate or label data, or who are affected by a machine learning system that uses the dataset.

These questions will help you select a dataset to work with, and to understand how its creation could inform your project if you use it. Often the answers to these questions can be found in a dataset's DATASHEET or a related research paper. Although including datasheets is becoming a standardized practice, not all datasets have datasheets, and many datasheets are incomplete. Look for datasets with complete datasheets, updated documentation, and current contact information for its creators.

### WHO CREATED THIS DATASET? WHO FUNDED IT? WHAT WERE THEIR MOTIVATIONS OR AIMS, AND HOW DO THEY COMPARE TO YOURS? [24]

If they differ in significant ways, consider how using the dataset for other purposes will impact both the original data subjects, as well as data subjectees, and the outcomes of your project. Would an alternative dataset be more appropriate? Document the rationale for the dataset you choose.

### HOW WAS THE DATA COLLECTED? HOW WAS IT ANNOTATED AND BY WHOM? ARE THE ORIGINAL ANNOTATION INSTRUCTIONS AVAILABLE? HOW ARE THE LIMITATIONS OF THOSE METHODS ACCOUNTED FOR? [47] WERE DATA SUBJECTS PART OF THE DATASET'S DESIGN AND CREATION? WAS THE RESULTING DATA VALIDATED BY ITS SUBJECTS? [24], [48]

If information about collection and annotation is missing, or if the collection and annotation methods were inappropriate or misaligned with your objectives, you may want to consider an alternative dataset. Also consider the contexts of labeling and annotation, as crowd-sourced data can lack the nuance of individual annotators from diverse perspectives who had the opportunity to collaborate [49].

HOW HAS THE DATASET BEEN PROCESSED ALREADY? IS IT A SMALL SAMPLE OF A LARGER COLLECTION? IS IT A COMPILATION OF OTHER PRE-EXISTING DATASETS? HAS IT BEEN STANDARDIZED OR TRANSFORMED IN ANY WAY (SEE SECTION 5)?

> If the dataset is a sample from a larger dataset or a compilation of smaller datasets, investigate those original sources to see if its data matches the sample or is more appropriate for your work. Does information in its datasheet impact your decision to use this dataset? If it has been transformed, see if the documentation also includes the original version or a description of its methods.

WHAT DOES THE DATASET CONTAIN? DOES IT INCLUDE A CODEBOOK DESCRIBING ITS PARTS? WHICH PERSPECTIVES ARE INCLUDED, AND WHICH ARE MISSING? WHICH OUTLIERS ARE DISMISSED, AND WHAT DATA IS UNACCOUNTED FOR? CAN YOU AUDIT THE DATASET, OR HAS IT ALREADY BEEN AUDITED? WHAT DOES THE AUDIT SHOW AND HOW CAN YOU ACCOUNT FOR ITS FINDINGS?

> If the dataset has gaps that neglect important considerations or that might affect your project, would a different dataset be more appropriate?
>
> Compare with other datasets in this area to see how they account for similar issues. Consult with your project group or community of practice to see how they interpret these issues and their importance. Get their help to spot issues you might have missed, and consider any other datasets that might better fit the project. Offer in-kind support.
>
> Regardless of whether you use the dataset, be sure to document your questions and concerns. Describe the limitations you see in the dataset, discuss their relevance to your project, and share what you have done to mitigate their impacts. Proceed with caution, if at all.

WHEN WAS THE DATASET MADE? IS THIS ITS LATEST VERSION? IF IT HAS BEEN DEPRECATED (OR REMOVED FROM PUBLIC CIRCULATION), WHY? DOES IT CONTAIN INFORMATION THAT IS INACCURATE OR OFFENSIVE? FOR MORE DISCUSSION OF DEPRECATED DATASETS, SEE SECTION 7.2.3.

> If the dataset is no longer valid, you will need to find another dataset. Maybe an updated version of the dataset addresses the issues that led to its removal, or perhaps you can make these revisions yourself. However, resolving ethics or accuracy issues is not as simple as updating a table, since the underlying structure of the dataset may remain problematic. Proceed with caution, if at all.

> Just because a dataset is *not* deprecated does not mean it is safe to use. There is not yet any standardized way to audit, update, or remove existing datasets, or even a central repository where issues can be documented and addressed. Much remains at the creators' discretion [50].

WHO IS FEATURED IN THE DATASET? WHO IS LEFT OUT? HOW DOES THE DATASET ACCOUNT FOR WHAT IS MISSING? WHAT ASSUMPTIONS, INTUITIONS, THEORIES, STEREOTYPES, OR INEQUITIES ARE CONTAINED IN THE DATA OR BUILT INTO THE DATASET'S STRUCTURE? HOW MIGHT THESE FRAMEWORKS HAVE BEEN PERPETUATED THROUGH FORMATTING AND TRANSFORMATION PROCESSES AS THE DATASET WAS MADE MACHINE-LEGIBLE?

> If you are unsure how the dataset and your use of it may impact a community, consult with its members to understand their perspectives. "Build with, not for," says the Design Justice Network, which emphasizes that "community members already know what they need and are working towards solutions that work for them" [51].
>
> Consider the consequences of inclusion. In an unjust system (e.g. racist criminal sentencing) more representation is not the right answer. Completeness may do more harm.
>
> Document your processes. Noticing what information is left out can be just as important as what is shown [48]. If too much information is missing, or omissions are too problematic, you may need to find another dataset.
>
> Maybe just don't build it. The best solution to a problem is not always more technology or a new system. Explaining why you did not build a model or did not use a dataset can still be a valuable contribution — showing the need for caution, skepticism, and alternative approaches.

HOW WAS CONSENT GIVEN FOR INCLUSION IN THE DATASET? WAS CONSENT FULLY INFORMED, VOLUNTARY, AND REVOCABLE? HOW ARE SUBJECTS' ANONYMITY PROTECTED?

> If the dataset does not adequately address how consent was provided, or if this consent does not extend to the uses of your project, you will need to find a different dataset. If your use potentially risks subjects' anonymity as established in the original dataset design, you will need to find a different dataset. Privacy cannot be guaranteed; be wary of the potential for re-identification of so-called anonymous data when they are combined with other datasets [52].
>
> If full consent was not given — especially if consent was not possible to obtain — "just don't build it" is always a valid and responsible option.

DO YOU HAVE LEGAL, ETHICAL ACCESS TO THIS DATASET? DOES YOUR USE OF IT ALIGN WITH ITS LICENSING AND TERMS OF USE, AND WITH YOUR OWN CODES OF CONDUCT? ARE ANY OF THE DATA INTENDED FOR RESTRICTED USE BY SPECIFIC COMMUNITIES? [53]-[55]

> We are not lawyers and this is not legal advice, of course. Check at the beginning, middle, and toward the end of your project — before publication or launch — that your use is authorized and appropriate to its creators and its subjects. It should be in keeping with both the letter and the spirit of terms of use and codes of conduct. Talk with others doing similar work to see what pitfalls they wish they had avoided upfront.

OVERALL, IS THIS THE BEST DATASET FOR YOUR PROJECT? WHAT ARE THE TRADEOFFS OF USING THIS DATASET VERSUS ANOTHER ONE?

# USAGE: WHAT IS THE STORY YOU WILL TELL WITH YOUR DATASET?

Your project aims will steer your dataset selection, the features you choose to interpret, the model you pick, and the adjustments you make. Across all of these small and large technical choices, you can use critical lenses to achieve your aims and minimize harms.

WHAT IS YOUR PROJECT'S GOAL? HOW DOES THE DATASET HELP YOU ACHIEVE IT?

> Prioritize your own purpose and approach over the popularity of the dataset or your familiarity with it. What is the best dataset for this task, to answer these questions?

WHAT ASSUMPTIONS ARE YOU PRIORITIZING OR EXCLUDING BY HOW YOU'VE FRAMED YOUR PROJECT? HOW MIGHT IT BE REFRAMED TO GAIN MORE INSIGHT? HOW MIGHT YOU COLLABORATE WITH PEOPLE FROM DIFFERENT DISCIPLINES OR BACKGROUNDS — E.G. ARTISTS, PRACTITIONERS, OR COMMUNITY STAKEHOLDERS — TO DEVELOP THE PROJECT'S AIMS MORE RICHLY?

Gather and prioritize the perspectives of data subjects and data subjectees. Connect with experts from other domains for fresh eyes and constructive feedback. Seek out information that you would not normally encounter; don't automatically rule out information in an unfamiliar vocabulary or format.

WHAT ASSUMPTIONS ARE YOU MAKING AS YOU PROCESS AND CLEAN THE DATASET, AND AS YOU SELECT FEATURES FOR ANALYSIS? COULD CATEGORIES (FEATURES) IN THIS DATASET BE DISAGGREGATED TO TELL A DIFFERENT STORY THROUGH THE DATA? SOME COMMON ASSUMPTIONS:

How have you treated the collected data as neutral in any obvious or more subtle ways?

Consider how transforming data [40] during the cleaning process may misrepresent information or remove important detail from the dataset. e.g. `NaN` (Not a Number) may conceal data that was never collected. (See Section 5 for more.)

A dataset is frequently treated as a generalizable, multi-purpose tool, applicable across many tasks and disciplines, when more often it is best suited or only suited to its original purpose.

Data cleaning is not one-and-done, but an iterative, integral process [42]. Though often undervalued labor, data cleaning is part of the 'real' work of making datasets.

WHAT ASPECTS OF THE DATASET WILL YOU INCLUDE OR EXCLUDE IN YOUR PROJECT, AND HOW DO THEY CONVEY THE INFORMATION? HOW WILL YOU ENSURE THESE CHOICES DO NOT OVERLY SKEW THE RESULTS?

FEATURE SELECTION & ENGINEERING. Distilling a dataset into pertinent columns is an essential part of dataset work because it determines what information categories will be important for later analysis. This process is descriptive and creative, not self-evident [26].
(See Section 3: FEATURE and Section 5: ENGINEERING FEATURES for more.)

#### HOW FAR DOES YOUR PROJECT DEVIATE FROM THE DATASET'S ORIGINAL PURPOSE? IF IT IS SIGNIFICANTLY DIFFERENT, HOW WILL YOU TRACK ANY NEW IMPACTS? [54]

When diverging from the original objectives of a dataset, ensure that your dataset transformation processes align with both your own goals and any guidelines put in place by the dataset's creators. Consider consent and licensing restrictions, as well as other potential legal and ethical issues. Furthermore, what kinds of impacts would the creators not have foreseen in your use of their dataset? How might communities be newly impacted by your use of this dataset, and how can you engage them in this process?

#### COULD YOUR USE OF THE DATASET CAUSE HARM? HOW WILL YOU MEASURE AND ADDRESS ADVERSE IMPACTS? WHAT STEPS WILL YOU TAKE TO MINIMIZE HARM? [54]

Work closely with data subjects and potential data subjectees who may be impacted by your use of the dataset, in order to discover what mitigation strategies would be best for them. How you address potential impacts will depend greatly on the types of risks your project presents; listening across diverse communities and disciplines will help you uncover new issues, create checkpoints, and take useful action. Encourage in your project team a culture of openness and learning from mistakes [48].

#### HOW WILL YOU MAINTAIN THE CONSENT AND ANONYMITY OF ANY DATA SUBJECTS? HAVE THEY BEEN TOLD ABOUT THE RISKS SPECIFIC TO YOUR USE CASE? CAN THEY CHECK BACK ON HOW THEIR DATA HAS BEEN USED?

Consent requires that subjects understand both the implications of data's use and impact, as well as their digital rights [48]. This goes beyond terms of service disclosures and should include the purpose of the project, any risks, and instructions on how to revoke consent if necessary.

There is no such thing as "anonymizing" identifying data because data can easily be combined with other sources to re-identify people [48].

DOES YOUR WORK WITH THIS DATASET RESULT IN A NEW, DERIVATIVE DATASET? HOW WILL YOU ACCOUNT FOR NEW ETHICAL CONCERNS ARISING FROM THE DERIVATIVE DATASET WHILE STILL ADDRESSING ISSUES RAISED BY THE ORIGINAL?

> Refer to Section 6.1 ORIGINS, as well as to resources for creating datasets like Gebru, et al.'s, "Datasheets for Datasets" to ensure that you are considering the questions that arise from creating a new or derivative dataset [24]. Your documentation, including a new datasheet, will help others who may want to use your new dataset.

OVERALL, HOW HAS YOUR INITIAL EXPLORATION OF THE DATASET CHANGED YOUR PROJECT OR ITS GOALS? WHAT NEW QUESTIONS DOES IT RAISE?

# STEWARDSHIP: WHAT STORY WILL THIS DATASET KEEP TELLING?

Although you may have completed your analysis, your work with the dataset is not done. Critical stewardship takes a holistic approach to sharing, maintaining, and deprecating a project. It's not just fixing something when it breaks, it requires sustainable and thoughtful relationships. Dataset stewardship lasts the whole data lifecycle.

HOW WILL YOU SHARE THIS DATASET? WILL YOU PROVIDE ACCESS TO YOUR MODIFIED VERSION OR LINK TO THE CREATORS' ORIGINAL VERSION? WHO WILL HAVE WHAT KINDS OF ACCESS (OPEN-SOURCE, AUTHENTICATED, PLATFORM-BASED)? [54]

> Maintaining any existing terms of service or consent agreements, consider making your dataset and results as available as possible. Keep in mind that the spirit of open access means more than just uploading your files to a repository or posting a link, it also includes sharing clear, complete documentation. Consider audiences outside your project team, including the data subjectees (see Section 3) and others who may be impacted, and include instructions for how to use the dataset in plain language. Code notebooks, examples, screenshots, and use cases are all helpful; and you can support broader access by using open-source, free tools, and multiple formats for creating your examples.

**ARE YOUR RESULTS FAIR (FINDABLE, ACCESSIBLE, INTEROPERABLE, REUSABLE)?** [56]

List your project findings and dataset with dataset repositories. Many repositories have a section to list multiple projects that cite a particular dataset. Make sure that your dataset has a DOI and complete metadata, and that files are in standard formats. Include a clear, comprehensive license so that it can be reused appropriately.[10]

**HOW WILL YOU DOCUMENT ANY ADDITIONAL DATA PREPROCESSING THAT WAS NECESSARY FOR YOUR USE OF THE DATASET? WHAT OTHER KINDS OF DOCUMENTATION (DATASHEETS, CODEBOOKS, ETC.) ARE NECESSARY?** [24]

Because you are using your dataset for a new project with a new objective, it makes sense to create a new datasheet. Cite the original datasheet and make any adjustments that reflect your project. E.g. if you created a new feature to study whether or not a user was answering a survey online, and this added a column to your version of the dataset, include that in the datasheet. Discuss why each decision was made and how the work was done. It may seem tedious, but keeping a research notebook or a working draft of your datasheet as you go can become a regular part of your practice, an easy way to document your work, and great help to other people who work with your dataset in the future.

**WHAT PRESENTATION FORMS WILL YOU USE TO TELL STORIES WITH THIS DATASET? CAN YOU COLLABORATE WITH OTHERS TO USE DIFFERENT FORMS OR ENGAGE DIFFERENT SENSES?** [2]

So much meaning is encoded in seemingly simple design choices. Make those choices intentionally, work with people who specialize in data-based storytelling, and value input and opportunities to reach new audiences with different forms. The Design Justice Network has a zine collection illustrating how to practice community-centered, equitable design [51].

**HOW WILL YOU MAINTAIN AND MONITOR ACCESS TO YOUR DATASET IN WAYS THAT CONSIDER THE INTELLECTUAL PROPERTY AND PRIVACY RIGHTS OF DATA SUBJECTS AND SUBJECTEES (SEE SECTION 3)?**

If your dataset has privacy or consent constraints, cultural considerations, proprietary constraints, or other reasons that it should not be shared broadly, make a plan for data storage. This plan should be as secure (if not more) and long-standing as the plan for the original dataset. Create a stewardship chain for the project and its infrastructure that will be maintained after you or your team have moved on [50].

### HOW WILL YOU KNOW IF THE DATASET'S CREATORS REVISE OR DEPRECATE THE DATASET, AND WHAT IS YOUR PLAN FOR HANDLING SUCH CHANGES? [50]

In your data stewardship plan, include regular checks of the original dataset's website or repository. Know how you will proceed if the dataset is revised or removed. This may mean revising your own dataset, revising your preprocessing, updating models, or reconfirming impacts to community members and their consent.

### WHO WILL ARCHIVE AND/OR DEPRECATE YOUR OWN DATA WHEN NECESSARY, AND HOW WILL THIS BE DONE?

Follow deprecation best practices using guidelines like the ones recommended by Luccioni and Corry, et al. [50] (discussed in Section 7). Their "Framework for Deprecating Datasets" suggests including the reasons for deprecation, how the removal will occur and plans for mitigating any negative impacts, an appeal mechanism for others who may be making use of your work, a timeline of the process, protocols for access after deprecation (usually with restrictions for research, legal, or historical use only), and a publication check request asking future paper authors to confirm they are not using a deprecated version.

**For your consideration:** Although principles like indigenous data governance or data feminism may seem abstract or hard to apply, they illustrate practices that can help you design better and more thoughtful projects that accomplish your goals and respect people who have historically been excluded from and harmed by dataset design.

| HOW DOES YOUR WORK ALIGN WITH PRINCIPLES OF DATA FEMINISM — I.E., EXAMINING AND CHALLENGING POWER, ELEVATING EMOTION AND EMBODIMENT, RETHINKING BINARIES AND HIERARCHIES, EMBRACING PLURALISM, CONSIDERING CONTEXT, AND VALUING LABOR? [47] |
|---|
| ARTICULATED IN THE CARE PRINCIPLES FOR INDIGENOUS DATA GOVERNANCE, HOW DOES THE DATASET OFFER COLLECTIVE BENEFIT; GIVE ITS SUBJECTS AUTHORITY TO CONTROL DATA; REQUIRE RESPONSIBILITY FROM PROJECT TEAMS; AND CENTER ETHICS, HUMAN RIGHTS, AND WELLBEING "AT ALL STAGES OF THE DATA LIFE CYCLE AND ACROSS THE DATA ECOSYSTEM"? [57] |

# CAUTIONS & REFLECTIONS FROM THE FIELD

How does mishandling datasets contribute to harm? Like any messy, multifaceted material, datasets must be treated with care. Taking time to see the broader implications of making and using datasets can save you time, create projects that are easier to explain, and help you build stronger relationships with the communities your datasets impact. Here we review why datasets matter to machine learning, why current approaches can sometimes be inadequate, and some lessons learned from those who have worked with machine learning datasets.

# DATASETS DIRECTLY IMPACT LIVES

Improper dataset use can impact DATA SUBJECTS (people contained in the original dataset), DATA SUBJECTEES (people analyzed or affected by a dataset's machine learning system, see Section 3), DATA WORKERS (the laborers who prepared the dataset), and even DATA RESEARCHERS, DESIGNERS, JOURNALISTS, ENGINEERS, ARTISTS, or other communities working with datasets. Whether the machine learning systems driven by datasets help deny resources to individuals (allocative harms), or misrepresent communities (representational harms) [58], they can powerfully and dangerously impact people.

Algorithmic processes often mirror and magnify existing power imbalances in social systems. In a process that sociologist Ruha Benjamin calls "coded inequity," algorithms and datasets "reflect and reproduce existing inequities" while simultaneously promoting them as "more objective or progressive." The speed and scale of machine learning and massive datasets make "discrimination easier, faster, and even harder to challenge" [59]. It also makes inequity more invisible and insidious, and dataset users can best understand the implications of their materials by looking carefully for impacts and working with others who can see impacts they might miss.

# WHERE DO DATASETS GO WRONG?

Whether designing a dataset from scratch or using one that has been around for years, decisions made at every step will inform your project outcomes. These decisions get scaled and compounded by machine learning models. This section summarizes common pitfalls of working with existing datasets and suggests ways to be cautious at various parts of a dataset project. In a survey of how machine learning researchers often work with large, complex datasets, natural language processing researcher Amandalynne Paullada, et al., found four kinds of COMMON DATASET PITFALLS (see box).

> **COMMON DATASET PITFALLS**
>
> **spurious tasks:** "where success is only possible [...] because the tasks themselves don't correspond to reasonable real-world correlations or capabilities"
>
> **artifacts in the data:** "which machine learning models can easily leverage to 'game' the tasks"
>
> **sloppy annotation or documentation:** a lack of reflective description can "erode the foundations of any scientific inquiry based on these datasets"
>
> **representation:** "wherein datasets are biased both in terms of which data subjects are predominantly included and whose gaze is represented" [60], {11}

Even those actively trying to fix datasets can experience these pitfalls. Paullada, et al. observed that datasets which were modified after their creation — often in attempts to improve a model's ability to generalize — were still susceptible to the same kinds of problems as the originals. They suggest "a broader view to be taken with respect to rethinking how we construct datasets for tasks overall," including dataset cultures around benchmarking, use and reuse, and licensing [60]. Importantly, they emphasize the need to move beyond technical fixes to consider dataset stewardship holistically, from project design to deprecation, from historical and technical foundations to field-wide approaches.

During dataset creation, classification thinking already shapes how data is collected, organized, and later perceived. In a multi-year analysis of outsourced data work, computer and information science researchers Milagros Miceli and Julian Posada found that the majority of classification choices for crowdworkers who are labeling data were simple 'Either-Or' selections, with little room for complexity. Workers were "encouraged to ignore ambiguity altogether" [62].{12} While binary thinking simplifies data collection and labeling, it codifies the viewpoints of those who create the binaries: "These examples of social classification and conceptualization are not just about cultural differences between requesters and data workers, but they reflect the prevalence of worldviews dictated by requesters and considered self-evident to them" [62]. It is critical to remember that crowdworkers' judgments and abilities to discern complexity are silenced in classification tasks that leave no room for debate or discussion, making the resulting classifications "cleaner" but far less valuable for appreciating the richness of data.

After datasets are created, they are often not audited, revised, or maintained. In many cases, they continue to be used despite having serious flaws. When datasets are reassessed, they might be deprecated for legal, organizational, social, or technical reasons. However, as machine learning researcher Sasha Luccioni and critical data scholar Frances Corry, et al., found (as part of the Knowing Machines project, from which this Guide also comes), frequently the processes of deprecation lacked the communication and transparency needed to dissuade usage effectively. They traced six popular datasets that continued to be used after they had been deprecated for privacy violations, offensive language and imagery, problematic descriptive categories, ethics board violations, and lack of consent uncovered by investigative journalism [50]. Sometimes dataset creators simply move on from their roles or the infrastructure does not exist to sustain the dataset. However, these "zombie datasets" continue to create problems when they keep circulating and feeding machine learning models. While continued use of datasets as training data for machine learning models is one danger, datasets may also persist when incorporated into other datasets [50]. Just because a dataset is available, do not assume that it is without problems, or that it has not been deprecated.

Datasets need ongoing stewardship and reassessment. They must be re-evaluated over time, due to changes in laws and across different jurisdictions [50]. Anyone who uses such datasets, even without knowing their problems, could be at risk, say Luccioni and Corry, et al., citing Google, Microsoft, and Amazon's legal repercussions for use of IBM's "Diversity in Faces" dataset [50]. Datasets also change context over time due to cultural changes (sometimes called "semantic drift") or repurposed uses that make their data irrelevant, inappropriate, or harmful.{13}

CAUTIONS & REFLECTIONS FROM THE FIELD

7.3

# WHY NOT SIMPLY 'DE-BIAS' DATASETS? BECAUSE "BIAS" IS ALWAYS BUILT-IN

Harms cannot be eliminated completely by removing problematic content, optimizing the system, or finding the perfect dataset. Although definitions of BIAS (both technical definitions and its varied cultural understandings) drive many conversations about accountability, diversity, fairness, ethics, explainability, and transparency in machine learning, "bias" is a complicated term. In a survey of almost 150 papers trying to address bias in machine learning, computer scientist Su Lin Blodgett, et al., found that authors struggled to reach consensus on definitions of "bias" and to articulate how the biased systems were harmful and to whom [66]. The term stands in for a complex set of concerns embedded more foundationally in machine learning systems, often centering on classification. Much interdisciplinary research examines classification — both its history and function as a fundamental mechanism of machine learning and also how its core principles operate technologically and socially — which this guide cannot fully address here.{14}

While important research in machine learning techniques is investigating how to create more robust models — whether by refining learning techniques, supplementing data with different or adversarial datasets, or applying other approaches [72]-[74], {15} — these quantitative strategies can patch key issues or improve accuracy, but they cannot fix underlying structural issues or embedded sociotechnical problems [75], [76]. Rather than attempting to remove bias or avoid classification altogether, work to move beyond bias-focused quick fixes. This guide recommends striving for a layered approach.

### AWARENESS:

First, understand the classifications in your dataset. What categories does it assume? Do research and work with your team to understand the power those categories have and the meanings they carry. Connect outside your circle to understand your dataset better.

ACTION:

Second, be prepared to shift your project if the classifications in your dataset are potentially too reductive, oppressive, or harmful — such that the dataset should not be used for machine learning. For situations in which the resulting system might contribute to power imbalances, or collapse complex identities or relations, just don't build it.

**For your consideration:** While it may be impossible to escape classification's worldviews entirely, with awareness of the underlying assumptions of classification and its impact on your processes, it becomes easier to make critical decisions that account for these contexts.

## INTERSECTIONAL APPROACHES TO DATASET PRACTICES

Many who work with datasets are already building alternative systems and strategies. Machine learning can be approached with fundamentally different mindsets and aims from the start. One approach being used to address this question is INTERSECTIONALITY, which is grounded in Black feminism and the legal theory of Kimberlé Crenshaw. Intersectionality analyzes how power operates at system-wide scales, sustaining oppression and shaping identities in layered ways. Intersectionality is also a set of active strategies developed by communities and passed down over time [43], [77]-[81]. Intersectional principles applied to datasets include centering those who have been at the margins and those impacted most, maintaining their priority at each phase of a dataset's lifecycle, and emphasizing ethics of relationality and care [81]-[83]. (For a variety of perspectives applying intersectionality to digital technologies, see the anthology The Intersectional Internet edited by internet researcher Safiya U. Noble and professor of education and psychology Brendesha M. Tynes [81] are some more strategies:

**Listen:** AI researcher Pratyusha Kalluri argues, "Researchers should listen to, amplify, cite and collaborate with communities that have borne the brunt of surveillance: often women, people who are Black, Indigenous, LGBT+, poor or disabled." Rather than focusing on fairness, they ask, "How is AI shifting power?" [84]

**Question assumptions, engage consequences:** Writer and data ethics researcher Anna Lauren Hoffman suggests that practitioners engage the "consequences of our work, but also our assumptions, our categories, and our position relative to the subjects of the data we work with" [85].

**Put power in context:** Information science researchers Miceli, Posada, and Yang recommend contextualized power-aware approaches that account for "historical inequities, labor conditions, and epistemological standpoints inscribed in data" [86].

**Strike a balance, share decisions:** Catherine Nicole Coleman suggests the information sciences have had to approach with a perspective of managing rather than eliminating classification and bias, by grappling with the dynamic balance between curating information and sharing it. It relies on such decisions being made over time and distributed among diverse groups [87].

**Ask essential questions:** Even before deciding whether an algorithm is the answer to a problem, technologist Kamal Sinclar recommends asking, "Can the available data lead to a good outcome?" and "Will the people affected by these decisions have any influence over the system?" [88] These are the kinds of questions this guide tries to parse out in detail for each stage of the dataset lifecycle.

# CONCLUSION

THIS GUIDE AIMS TO HELP YOU WORK CRITICALLY WITH MACHINE LEARNING DATASETS:

- to see existing datasets from different perspectives;
- to appreciate the complexities in their origins, classifications, and transformations;
- to read the messiness across the lifecycles of datasets;
- to reach out to people impacted by your dataset work;
- and, perhaps most importantly, to understand the benefits of advancing your projects with thoughtful, accountable data stewardship.

We hope that you dip into it, revisit parts, follow-up on references, and share pieces that you think are helpful to your teams and communities. Given the recent proliferation of text-to-image machine learning and synthetic data, and no doubt new tools and applications all the time, we hope you also use the critical perspectives and guidelines you develop here with new technologies as they emerge. We intended this guide not as a definitive source — many of these topics are too big to be covered completely — but as a starting point for your explorations and an illustration of how productive and fruitful it can be to approach dataset work critically.

We also hope the guide is a prompt for more interdisciplinary and intersectional conversations about critical dataset work — and the promise of conscientious approaches to machine learning. With continued efforts, our hope is for this field guide to be a living document with expansions, updates, and additional resources as the dynamic world of machine learning datasets continues to evolve.

# CREDITS, ACKNOWLEDGMENTS

## CREDITS

Author: Sarah Ciston

Editors: Mike Ananny and Kate Crawford

Design and illustrations: Vladan Joler and Olivia Solis

Published by: Knowing Machines project (https://knowingmachines.org)

Full citation: S. Ciston, "A CRITICAL FIELD GUIDE FOR WORKING WITH MACHINE LEARNING DATASETS," K. Crawford and M. Ananny, Eds., Knowing Machines project, Feb. 2023.

We wish to thank all the members of the Knowing Machines research project, including Christo Buschek, Franny Corry, Melodi Dincer, Vladan Joler, Ed Kang, Sasha Luccioni, Will Orr, Jason Schultz, Hamsini Sridharan, and Jer Thorp for their fruitful conversations and generous feedback on drafts of this work, and Hannah Franklin and Michael Weinberg for their administrative support. Warm thanks go to Lee Kezar for their rigorous technical perspective and contributions, and also much appreciation to Will Orr for proofreading, Vladan Joler and Olivia Solis for design, and Michael Weinberg for project management. We also want to acknowledge the members of the USC Libraries' "Visual Datasets for Inclusive Research" project, including Hujefa Ali, Bill Dotson, Curtis Fletcher, Mike Jones, Caroline Muglia, and Manasa Rajesh for providing inspiration for this work, valuable perspectives on library collections as data, and a warm and enriching research environment. We want to acknowledge the support of the Alfred. P Sloan Foundation, as part of their funding of the Knowing Machines project. Finally, the author wishes to thank editors Kate Crawford and Mike Ananny for their consistently kind and thoughtful support throughout.

## DESIGN NOTE

In the tradition of the early net.art experimentation, this Guide was entirely created within a spreadsheet. This experimental design concept is exploring possibilities and constraints of the spreadsheet as a medium that plays an important role in the creation of the machine learning datasets. The illustrations, inspired by early modernism and optical art, play with the idea of a "bureaucratic modernism"-style of art, fitting for an age where everyone is expected to take on the roles of both manager and bureaucrat.

## ABOUT KNOWING MACHINES

This critical field guide is published by Knowing Machines. Knowing Machines is a research project tracing the histories, practices, and politics of how machine learning systems are trained to interpret the world.

Our group develops methodologies and tools for understanding, analyzing, and investigating training datasets, and studying their role in the construction of "ground truth" for machine learning. We research how datasets index the world, make predictions, and structure knowledge cultures. We are an international team, and we aim to support the emerging field of critical data studies by contributing original research, reading lists, research tools, and supporting communities of inquiry that address the foundational epistemologies of machine learning. Knowing Machines is sponsored by the Alfred P. Sloan Foundation.

# ENDNOTES

{1} ciston@usc.edu, University of Southern California

{2} ananny@usc.edu, University of Southern California

{3} kate.crawford@usc.edu, University of Southern California, MSR-NYC

| {4} | This field guide encourages a combination of criticality and care toward datasets — plus machine learning materials and processes more widely, and the people and environments they impact — so that the meaning of 'data stewardship' resonates with its connotations of environmental stewardship and conservation, as well as its definitions of technical responsibility. Media scholar Yanni Alexander Loukissas lays out this approach in his book All Data Are Local. He says, "I take a critical stance, but also explore approaches to working with data that are less distant and cerebral than critical reflection implies. In order to do so, my approach integrates lessons from the feminist ethics of care. [...] Care is critical in that it calls attention to neglected things. But it is more than critical reflection; it is a doing practice" [1]. As Catherine D'Ignazio and Lauren Klein suggest in Data Feminism, this approach emphasizes embodiment and material contexts, with the potential to rethink hierarchies and challenge power [2]. |
|---|---|
| {5} | There is a growing body of scholarship emphasizing the need to approach datasets critically. For recent research on this topic, see cultural media theorist Nanna Bonde Thylstrup's introduction to "critical dataset studies" [3] and the dataset accountability framework and matrix of dataset development harms from Yale Law resident fellow Mehtab Khan's and Distributed AI Research Institute research director Alex Hanna [94]. Find more resources on the "Critical Dataset Studies Reading List" compiled by the Knowing Machines research project. |
| {6} | That the language of big data is reminiscent of global colonialist exploitation ("scrape," "extract," "capture," "the new oil") should be telling. Work by critical surveillance scholar Simone Browne and by Nick Couldry and Ulises A. Mejias, among others, has already traced the colonialist legacies big data builds upon [4], [5]. |
| {7} | For more on the sociotechnical phenomenon, see CLASSIFICATION THINKING. For more on machine learning definitions of classification tasks, as well as a technical introduction to data science concepts, see Computational and Inferential Thinking, edited by statistics professor Ani Adhikari, et al. [7]. |
| {8} | 'Data' and 'information' are complex terms. There is a healthy scholarly conversation about the many meanings of each word, beyond the scope of this guide, including Tarleton Gillespie's "The Relevance of Algorithms" [9]; Lisa Gitelman's edited volume "Raw Data" Is an Oxymoron [10], Colin Koopman's How We Became Our Data [11]; Christine L. Borgman's Big Data, Little Data, No Data [12]; and Rob Kitchin's Data Lives [13]. For more on information, see Ann Blair, et al.,'s Information: A Historical Companion and James Gleik's The Information: A History, A Theory, A Flood [14]; and for a detailed description of additional terms, see the anthology Uncertain Archives: Critical Keywords for Big Data [15]. |
| {9} | TOOLS OF THE TRADE: If you're working in the PYTHON programming language, you might use two popular tools called NUMPY and PANDAS to make these adjustments. They are both LIBRARIES or MODULES, which are add-on packs of software available to import. Often created with a specific field or task in mind, libraries are written on top of Python so that you don't have to write everything you want to do from scratch. Numpy helps with handling large groups of numbers, and Pandas has built-in support for lots of data manipulation tasks. You may also encounter the MATPLOTLIB library for making visualizations and SCIKIT-LEARN, KERAS, or other machine learning libraries down the line. |
| {10} | Popular dataset repositories include Hugging Face, Kaggle, Papers with Code, the Registry of Research Data Repositories, and Zenodo [89]–[93]. |
| {11} | MORE PITFALLS: MIT researchers Harini Suresh and John Guttag break data representation down further into seven types of harm, which they refer to as "bias," encountered across machine learning processes. These include historical: e.g. word embeddings that reflect and reinforce stereotypes; representational: e.g. underrepresenting or misrepresenting the target group, through limited sampling or a mismatch between target and use populations; measurement: e.g. variations in accuracy or method across groups; learning: e.g. pruning the data to enhance performance ends up amplifying disparities on underrepresented characteristics; evaluation: e.g. comparison against standardized benchmarks fails to detect issues when the benchmarks themselves are also biased; aggregation: e.g. applying an overgeneralized assumption to an entire set when subsets should be represented differently; and deployment: e.g. misalignment of how a dataset or model was designed and how it is used in practice [61]. |
| {12} | EXPLAINING & DESCRIBING, OR EXTRACTING & PRESCRIBING: Prediction, says communications scholar Sun-Ha Hong, "sees what it knows to see, and it measures what it can typically imagine measuring. These tendencies are shaped through longstanding economic and political asymmetries, whose influence is regularly written off as uncertain and uncontrollable 'externalities'. [...T]hey obfuscate how patterns of extraction shape the research questions and the choice of what to measure (and what to dismiss without measuring)" [63]. In their work on data colonialism, Nick Couldry and Ulises A. Mejias show that digital datasets join a much longer history of extraction [5]. As digital media researcher Wendy H.K. Chun argues, the materials and methods of machine learning — including datasets — work by forecasting the future from past data and prescribing what they purport to describe [64]. |
| {13} | For a thoughtful discussion of how datasets become recontextualized — and in particular the importance of using critical and historical analyses to question the continued circulation of datasets in communities not accustomed to their original contexts — see research on dataset audits by critical technology researcher Os Keyes and librarian Jeanie Austin [65]. |
| {14} | CLASSIFICATION THINKING: For multifaceted perspectives on the logics and politics of classification, please see, among many others: In Sorting Things Out [67], informatics professor Geoffrey C. Bowker and sociologist Susan C. Starr suggest that even if categories often feel invisible, "The material force of categories appears always and instantly." Creating a category draws a boundary and fixes the concept of what is inside and outside, as seen from the perspective of whomever has the power to create it. "Categories simplify and freeze nuanced and complex narratives, obscuring political and moral reasoning behind a category," argue computer scientists Vinay Uday Prabhu and Abeba Birhane [68]. Library scholar Hope A. Olson points out that classification problems are not new to the machine learning field; rather, the problematic goal to find "an overriding unity in language" has been codified through library catalog practices since at least the nineteenth century. It echoes Enlightenment-era impulses to "know" the world comprehensively [69], [70]. Furthermore, algorithmic attempts to understand people through classification are drawing on much longer practices of human exploitation that have created and justified categories of difference. In Dark Matters, critical surveillance scholar Simone Browne traces data practices like surveillance and biometrics back to the documents of the transatlantic slave trade, arguing that, "human categorization and division is part of a larger imperial project of colonial expansion that aimed to fix, frame, and naturalize discursively constructed difference" [4]. These are just a few (non-exhaustive) touch points for the discussion around classification thinking and machine learning. For more, you can look to the "Critical Dataset Studies Reading List," compiled by the Knowing Machines research project [22], or the "Critical Algorithm Studies: a Reading List" curated by Tarleton Gillespie and Nick Seaver [71]. |
| {15} | A meta analysis by Dieuwke Hupke et al. found that a majority (66%) of efforts focused on "practical" improvements, while only 2.6% focus on fairness. This included generalizability, considering in what kinds of situations a model can be applied: "One question that is often addressed with a primarily practical motivation is how well models generalise to different domains or differently collected data." Meanwhile, fairness research "asks questions about how well models generalise to diverse demographics, typically considering minority or marginalised groups [...] or investigates to what extent models perpetuate (undesirable) biases learned from their training data" [75]. |

A CRITICAL FIELD GUIDE FOR WORKING WITH MACHINE LEARNING DATASETS